\documentclass[12pt]{article}
\usepackage{graphicx}

\newcommand{\beq}{\begin{equation}}
\newcommand{\eeq}{\end{equation}}
\newcommand{\beqn}{\begin{eqnarray}}
\newcommand{\eeqn}{\end{eqnarray}}
\newcommand{\beqns}{\begin{eqnarray*}}
\newcommand{\eeqns}{\end{eqnarray*}}

\begin{document}

\begin{titlepage}
\begin{center}

\hfill USTC-ICTS-14-14\\
\hfill August 2014

\vspace{2.5cm}

{\large {\bf  Analysis of $J/\psi(\eta_c)\to \gamma ~+$ invisible decays in the standard model}}\\
\vspace*{1.0cm}
 {Dao-Neng Gao$^\dagger$ \vspace*{0.3cm} \\
{\it\small Interdisciplinary Center for Theoretical Study,
University of Science and Technology of China, Hefei, Anhui 230026
China}}

\vspace*{1cm}
\end{center}
\begin{abstract}
\noindent
Motivated by the experimental study of $J/\psi\to\gamma ~+$ invisible decay by the CLEO Collaboration, we analyze the process $J/\psi\to \gamma \nu\bar{\nu}$, as the standard model background for this invisible decay, at the lowest order. Our investigation shows that  Br($J/\psi\to \gamma \nu\bar{\nu}$) is far below the present experimental upper limit on the branching ratio of $J/\psi\to\gamma ~+$ invisible, which indicates that some interesting room for new physics in this channel might be expected. We also present a similar analysis of the $\eta_c\to \gamma \nu\bar{\nu}$ decay.
\end{abstract}

\vfill
\noindent
$^{\dagger}$ E-mail:~gaodn@ustc.edu.cn
\end{titlepage}
Charmonium radiative decay has long been an interesting topic for theoretical and experimental investigations.  The decay mode $J/\psi \to \gamma + X$, where $X$ is a narrow state that is invisible to the detector, has been analyzed by the CLEO Collaboration \cite{CLEO2010}, and the upper limit on the branching ratio has been reported, which could reach $6.3\times 10^{-6}$ at the 90\% confidence level.  It is generally thought that the invisible particle $X$ could be the neutrino in the standard model, and also the new one outside the standard model, which may be the light candidate of dark matter or other exotic particles from some new physics scenarios. Theoretically, in Ref. \cite{BM2000}, $J/\psi$ decays into a photon and Kaluza-Klein excitations of the graviton, induced by large extra dimensions formalism, have been discussed and a branching ratio around $10^{-5}$ may be expected. Thus, studies of quarkonium radiative decay to invisible final states may help us to explore the novel dynamics or provide some useful constraints on some models beyond the standard model.

Motivated by the experimental and theoretical efforts mentioned above, in this paper we will compute the standard model contribution to $J/\psi\to \gamma ~+$ invisible; namely, we shall focus on the analysis of the decay $J/\psi\to \gamma \nu\bar{\nu}$ since only neutrinos are invisible particles in the standard model. Only after we fully understand its standard model background can the future precise experimental investigations of the $J/\psi\to \gamma ~+$ invisible decay possibly provide us with some interesting information on new physics. To our knowledge, this analysis has not been done yet. We will also present a similar analysis of $\eta_c\to \gamma \nu\bar{\nu}$ decay.

\begin{figure}[t]
\begin{center}
\includegraphics[width=12cm,height=3cm]{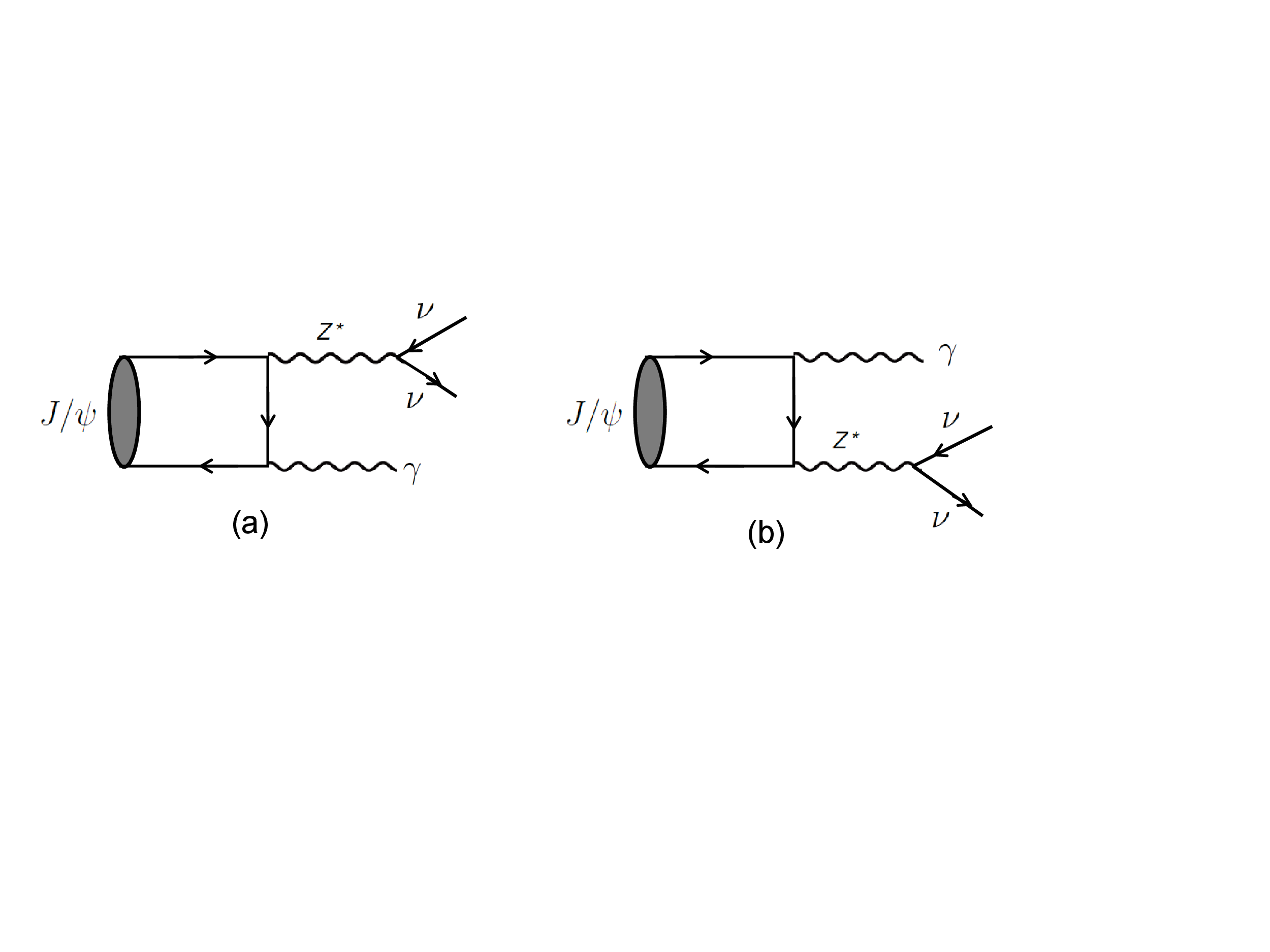}
\end{center}
\caption{Lowest-order diagrams for the decay $J/\psi\to \gamma \nu\bar{\nu}$.}\label{figure1}
\end{figure}

We consider the decay in the rest frame of $J/\psi$,
\beq J/\psi (P)\to \gamma (q)+\nu (p_1) + \bar{\nu} (p_2),\eeq
where momenta are given in brackets. In the standard model, the lowest-order contribution to the process is from the transition $J/\psi\to \gamma Z^*$ with $Z^*$ converted into the neutrino pair, which has been shown in Fig. \ref{figure1}. Vertices in these diagrams are given by the neutral current interactions guiding dynamics of the photon and $Z$ boson couplings to fermions, which can be expressed as
\beq\label{NC}
{\cal L}_{\rm NC}=e J_\mu^{\gamma} A^\mu+\frac{g}{2\cos\theta_W}J^Z_\mu Z^\mu\eeq
with
\beq\label{emcurrent}
J_\mu^{\gamma}=\sum_f Q_f\bar{f}\gamma_\mu f,
\eeq
and
\beq\label{weakneutralcurrent}
J_\mu^Z=\sum_f \bar{f}\gamma_\mu(g_V^f-g_A^f\gamma_5)f,\eeq
 where $e$ is the coupling constant of electromagnetic interaction, $g$ is the SU(2)$_L$ coupling constant, $\theta_W$ is the Weinberg angle, and $f$ denotes fermions including leptons and quarks.  Also $g_V^f=T_3^f-2 Q_f \sin^2 \theta_W$ and $g_A^f=T_3^f$,  where $Q_f$ is the charge, and $T_3^f$ is the third component of the weak isospin of the fermion. For the charm quark and the neutrino, we know $g_A^c=g_A^\nu=g_V^\nu=1/2$, $g_V^c=1/2-4/3\sin^2\theta_W$, and $Q_c=2/3$.

On the other hand, one can easily find that the diagrams in Fig. \ref{figure1} actually depict the partonic level transition $c (p_c)\bar{c}(p_{\bar{c}})\to \gamma (q) + Z^* (k)$ with $k=p_1+p_2$. In order to obtain their contributions to the hadronic decay amplitude, one has to project the quark pair into the corresponding hadron state. For $J/\psi$, the only one hadron in the process, a good approximation is to use the nonrelativistic color-singlet model \cite{colorsinglet}.  In this model the quark momentum and mass are taken to be one half of the corresponding quarkonium momentum and mass, which means $p_c=p_{\bar{c}}=P/2$ and $m_{J/\psi}=2m_c$.   Thus, for the $c\bar{c}$ pair to form $J/\psi$, it should be projected into a vector color-singlet state,  and one can replace the combination of the Dirac and color spinors for $c$ and $\bar{c}$ by the following projection operator \cite{barger, Jia2007}:
\beq\label{projector}
u(p_c)_i\bar{v}(p_{\bar{c}})_j\longrightarrow \frac{\psi_{J/\psi}(0)I_c}{\sqrt{12m_{J/\psi}}}\left((P\!\!\!\!/+m_{J/\psi}){\epsilon\!\!/}_{J/\psi}\right)_{ij}~ ,\eeq
where $\epsilon_{J/\psi}^\mu$ is the polarization vector of $J/\psi$, and $I_c$ is the $3\times 3$ unit matrix in color space. $\psi_{J/\psi}(0)$ is the wave function at the origin for the $J/\psi$ state, which is a nonperturbative parameter. Likewise, for the $\eta_c$ state, one can adopt the projector as follows:
\beq\label{projectoretac}
u(p_c)_i\bar{v}(p_{\bar{c}})_j\longrightarrow \frac{\psi_{\eta_c}(0) I_c}{\sqrt{12 m_{\eta_c}}}\left((P\!\!\!\!/+m_{\eta_c})i\gamma_5\right)_{ij}~ .
\eeq

\begin{figure}[t]
\begin{center}
\includegraphics[width=11cm,height=8cm]{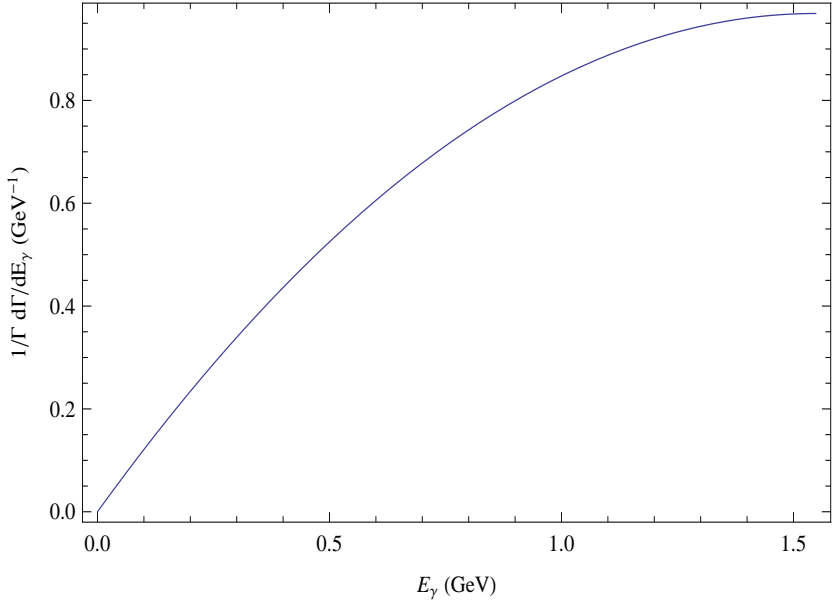}
\end{center}
\caption{The normalized decay spectrum for $J/\psi \to \gamma\nu\bar{\nu}$.}\label{figure2}
\end{figure}

Now let us commence with the decay amplitude of $J/\psi\to \gamma\nu\bar{\nu}$. Direct calculation, by employing eqs. (\ref{projector}), (\ref{NC}), (\ref{emcurrent}) and (\ref{weakneutralcurrent}), will lead to
\beqn\label{amplitude1}
{\cal M}(J/\psi\to \gamma \nu\bar{\nu})&=&-\frac{4 \sqrt{3}e Q_c g_A^cg_A^\nu G_F}{\sqrt{2}}\frac{\sqrt{m_{J/\psi}}\psi_{J/\psi}(0)}{m_{J/\psi}^2-k^2}\nonumber\\
&&\times~\varepsilon_{\alpha\beta\mu\nu}\epsilon_{J/\psi}^\alpha(P)\epsilon^{*\nu}(q) q^\beta \bar{u}(p_1)\gamma^\mu(1-\gamma_5)v(p_2),\eeqn
where $\epsilon^{*\nu}(q)$ is the polarization vector of the photon. Note that the momentum square of the virtual $Z$ boson $k^2\ll m_Z^2$, we have made an approximation by the replacement
\beq\label{approximation}\frac{1}{k^2-m_Z^2}\longrightarrow -\frac{1}{m_Z^2}
\eeq
for the $Z$ propagator, and
\beq\label{GF}\frac{G_F}{\sqrt{2}}=\frac{g^2}{8m_W^2}=\frac{g^2}{8 m_Z^2\cos^2\theta_W}\eeq
has been used in the derivation. Since charge conjugate invariance requires the axial coupling between the $Z$ boson and charm quark in Fig. \ref{figure1}, only $g_A^c$ appears in the amplitude (\ref{amplitude1}). One can check the Appendix for some details in deriving of the amplitude. Thus the differential decay rate of $J/\psi\to \gamma \nu\bar{\nu}$, in terms of $E_\gamma$, the photon energy in the rest frame of $J/\psi$,  can be written as
\beq\label{rate1}\frac{d\Gamma}{d E_\gamma}=\frac{4Q_c^2\alpha_{\rm em} G_F^2}{3\pi^2}\frac{|\psi_{J/\psi}(0)|^2}{m_{J/\psi}} E_\gamma(m_{J/\psi}-E_\gamma),
\eeq
where $\alpha_{\rm em}=e^2/4\pi$, and  the range of the variable is given by \beq 0\leq E_\gamma\leq m_{J/\psi}/2. \eeq
In deriving eq. (\ref{rate1}), we have taken the relation $k^2=m^2_{J/\psi}-2 m_{J/\psi} E_\gamma$, $g_A^c=g_A^\nu=1/2$, and the tiny neutrino mass has been set to zero. The decay spectrum given by  eq. (\ref{rate1}) has been plotted, as the function of the photon energy $E_\gamma$, in Fig. \ref{figure2}. The advantage of plotting the distribution is that the nonperturbative parameter $\psi_{J/\psi}(0)$ naturally cancels out in the normalized decay spectrum. One can find that it will get the maximum in the spectrum when $E_\gamma$ is taken to be half of the $J/\psi$ mass.

\begin{figure}[t]
\begin{center}
\includegraphics[width=7cm,height=2cm]{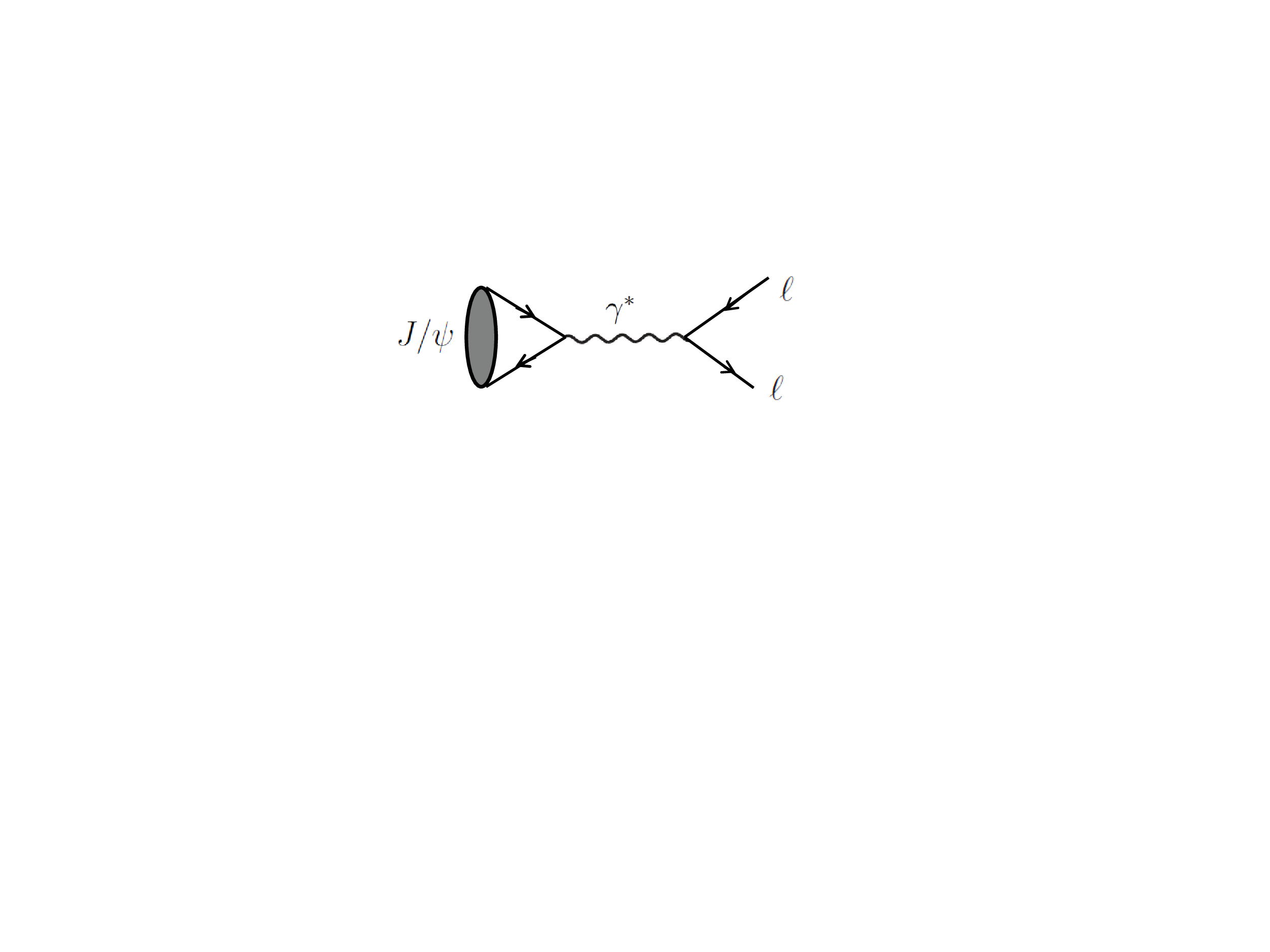}
\end{center}
\caption{Lowest-order diagram for the decay $J/\psi\to \ell^+\ell^-$.}\label{figure3}
\end{figure}

After integrating over $E_\gamma$ in the differential rate of  (\ref{rate1}), one can get the desired partial width as  \beq\label{width1}
\Gamma(J/\psi\to \gamma \nu\bar{\nu})=\frac{Q_c^2\alpha_{\rm em} G_F^2 m_{J/\psi}^2}{3\pi^2}|\psi_{J/\psi}(0)|^2,\eeq
in which three flavors of neutrinos have been summed. As the lowest-order evaluation, it is conventional to introduce the normalized decay rate of $J/\psi\to\gamma\nu\bar{\nu}$, defined as
\beq\label{normalizedrate} R(J/\psi\to\gamma\nu\bar{\nu})\equiv\frac{\Gamma(J/\psi\to\gamma\nu\bar{\nu})}{\Gamma(J/\psi\to \ell^+\ell^-)},
\eeq
normalized to the partial width of $J/\psi$ decaying into a charged lepton pair ($e^+e^-$ or $\mu^+\mu^-$). The lowest-order contribution to this decay rate, as depicted in Fig. \ref{figure3}, can be similarly computed, which is given by
\beq\label{width2}\Gamma(J/\psi\to \ell^+\ell^-)=\frac{16\pi Q_c^2\alpha_{\rm em}^2}{m_{J/\psi}^2}|\psi_{J/\psi(0)}|^2.
\eeq
This leads to
\beq\label{ratio1}
R(J/\psi\to \gamma\nu\bar{\nu})=\frac{G_F^2m_{J/\psi}^4}{48\pi^3 \alpha_{\rm em}}=1.2\times 10^{-9}.\eeq
Using the experimental data of  Br$(J/\psi\to \ell^+\ell^-)=(5.94\pm 0.06)\%$ for $\ell=e$ or $\mu$ in Ref. \cite{PDG2012}, we have
\beq\label{br1}  {\rm Br}(J/\psi\to \gamma\nu\bar{\nu}) = 0.7\times 10^{-10},
\eeq
  which is quite small, comparing with present upper limit on the branching ratio of $6.3\times 10^{-6}$, reported by the CLEO Collaboration \cite{CLEO2010, PDG2012}. This indicates that some interesting room for new physics in the decay $J/\psi\to \gamma ~+$ invisible might be expected. Although eq. (\ref{ratio1})is from the lowest-order calculation, it is reasonable to assume that high-order corrections would not change it very greatly, at least for its order of magnitude.

\begin{figure}[t]
\begin{center}
\includegraphics[width=5cm,height=2cm]{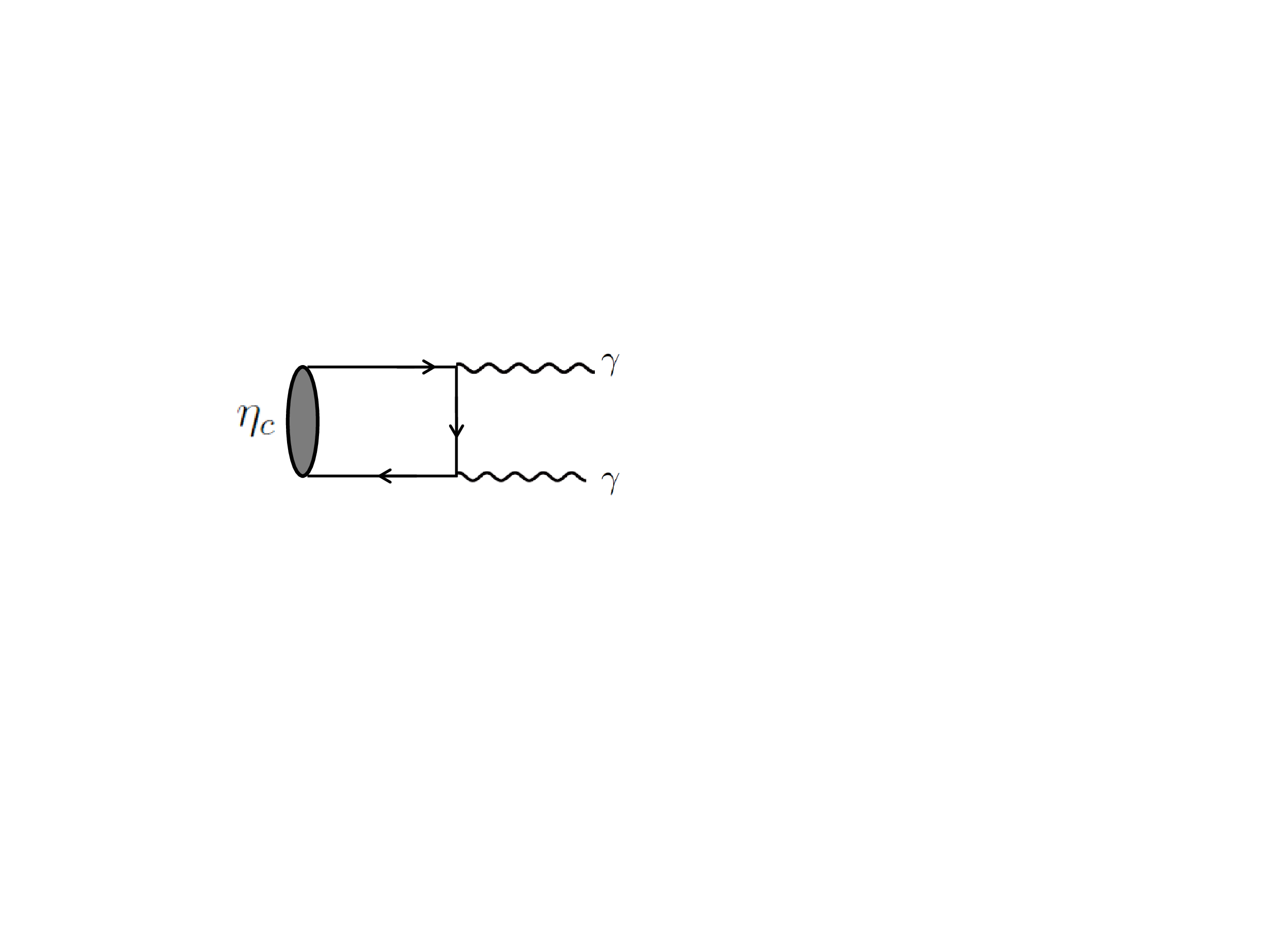}
\end{center}
\caption{Lowest-order diagram for the decay $\eta_c\to \gamma\gamma$.}\label{figure4}
\end{figure}

Next let us deal with pseudoscalar charmonium decay $\eta_c (P) \to\gamma(q)\nu(p_1)\bar{\nu}(p_2)$. In the standard model, the lowest-order contribution to the decay is also from the transition $\eta_c\to\gamma Z^*\to\gamma \nu\bar{\nu}$, and the corresponding diagrams are the same as Fig. \ref{figure1}. Analogously to the $J/\psi$ case and employing the projector (\ref{projectoretac}) for $\eta_c$, one can obtain the decay amplitude
\beq\label{ampetac}{\cal M}(\eta_c\to\gamma\nu\bar{\nu})=\frac{-4\sqrt{3}e Q_c g_V^c g_V^\nu G_F}{\sqrt{2m_{\eta_c}}}\frac{\psi_{\eta_c}(0)}{m^2_{\eta_c}-k^2} \varepsilon_{\alpha\beta\mu\nu}P^\alpha\epsilon^{*\nu}(q) q^\beta \bar{u}(p_1)\gamma^\mu(1-\gamma_5)v(p_2),
\eeq
where $k=p_1+p_2$ denotes the momentum of the virtual $Z$ boson. Straightforwardly, we get the partial width of the process
\beq\label{widthetac}
\Gamma(\eta_c\to\gamma\nu\bar{\nu})=\frac{Q_c^2\alpha_{\rm em} G_F^2 m^2_{\eta_c}}{4\pi^2}\left(1-\frac{8}{3}\sin^2\theta_W\right)^2 |\psi_{\eta_c}(0)|^2 ,
\eeq
where $g_V^\nu=1/2$ and $g_V^c=1/2-4/3 \sin^2\theta_W$ have been taken. Similarly, in order to get rid of the nonperturbative factor $\psi_{\eta_c}(0)$, we define the normalized decay rate as
\beq\label{ratioetac}
R(\eta_c\to\gamma\nu\bar{\nu})=\frac{\Gamma(\eta_c\to\gamma\nu\bar{\nu})}{\Gamma(\eta_c\to\gamma\gamma)}. \eeq
It is easy to evaluate the lowest-order contribution to the decay rate of $\eta_c\to\gamma\gamma$, as depicted in Fig. \ref{figure4}, which reads
\beq\label{etagammagamma} \Gamma(\eta_c\to\gamma\gamma)=\frac{48\pi Q_c^4 \alpha_{\rm em}^2 }{m_{\eta_c}^2}|\psi_{\eta_c}(0)|^2.
\eeq
Consequently,
\beq\label{ratioetac2}
R(\eta_c\to\gamma\nu\bar{\nu})=\frac{G_F^2 m_{\eta_c}^4}{192\pi^3\alpha_{\rm em}Q_c^2}\left(1-\frac{8}{3}\sin^2\theta_W\right)^2=8.1\times 10^{-11}.\eeq
Taking the data Br$(\eta_c\to\gamma\gamma)=(1.57\pm 0.12)\times 10^{-4}$ \cite{PDG2012}, we obtain
\beq\label{br2}{\rm Br}(\eta_c\to\gamma\nu\bar{\nu})=1.3\times 10^{-14}.\eeq
This is really a very small branching ratio.

In conclusion, charmonium radiative decay $J/\psi\to\gamma\nu\bar{\nu}$, which is the standard model contribution to the $J/\psi\to \gamma ~+$ invisible decay, has been computed at the lowest order. The nonrelativistic color-singlet model has been employed for the charmonium hadron state. Our study shows that Br$(J/\psi\to\gamma\nu\bar{\nu})$ is $0.7\times 10^{-10}$, which is far below the present upper limit on the branching ratio of $J/\psi\to \gamma ~+ $ invisible, reported by the CLEO Collaboration. This means that substantial room for new physics may exist in this process. Therefore, in future precise experiments, such as the super tau-charm factory, the $J/\psi\to \gamma ~+ $ invisible decay could be an interesting channel in searching for new physics scenarios beyond the standard model. We also carry out a similar analysis of the mode $\eta_c\to \gamma\nu\bar{\nu}$, as the standard model background for the $\eta_c\to \gamma ~+$ invisible decay. It is found that its branching ratio is strongly suppressed.

\vspace{0.5cm}
\section*{Acknowledgements}
This work was supported in part by the NSF of China under Grants No. 11075149 and 11235010.

\appendix
\newcounter{pla}
\renewcommand{\thesection}{\Alph{pla}}
\renewcommand{\theequation}{\Alph{pla}\arabic{equation}}
\setcounter{pla}{1}
\setcounter{equation}{0}

\section*{Appendix: Derivation of the decay amplitudes }

First the contribution to the partonic level transition $c (P/2)\bar{c}(P/2)\to \gamma (q) + Z^* (k) $ from Figs. 1(a) and 1(b) can be explicitly written as
\beqn\label{a1} && \bar{v}(P/2)\left[\gamma^\mu(g_V^c-g_A^c\gamma_5)\frac{P\!\!\!\!//2-q\!\!\!/+m_c}{(P/2-q)^2-m_c^2}\gamma^\nu+\gamma^\nu \frac{q\!\!\!/-P\!\!\!\!//2+m_c}{(q-P/2)^2-m_c^2}\gamma^\mu(g_V^c-g_A^c\gamma_5) \right]u(P/2)\nonumber\\
&& ~~~~\times \epsilon^*_\nu (q) \epsilon_\mu^{Z^*} \frac{-ie g Q_c}{2 \cos\theta_W},
\eeqn
where $\epsilon_\mu^{Z^*}$ denotes the virtual $Z$ boson, and  $p_c=p_{\bar{c}}=P/2$ has been set. Thus, employing Eq. (\ref{projector}), one will further get
\beqn\label{a2}
\frac{-i e g Q_c}{2\cos\theta_W}\frac{-2}{m_{J/\psi}^2-k^2}\frac{\sqrt{3}\psi_{J/\psi}(0)}{2\sqrt{m_{J/\psi}}}\cdot {\rm tr}(P\!\!\!\!/+m_{J/\psi}){\epsilon\!\!/}_{J/\psi}\left[\gamma^\mu(g_V^c-g_A^c\gamma_5)(P\!\!\!\!//2-q\!\!\!/+m_c)\gamma^\nu \right.\nonumber\\
\left. +\gamma^\nu(q\!\!\!/-P\!\!\!\!//2+m_c)\gamma^\mu(g_V^c-g_A^c\gamma_5)\right]\epsilon^*_\nu (q)\epsilon_\mu^{Z^*}
\eeqn
for the $J/\psi\to\gamma Z^*$ transition. Here we have taken $m_c=m_{J/\psi}/2$ and $(P/2-q)^2-m_c^2=-(m_{J/\psi}^2-k^2)/2$, and tr denotes the trace over the Dirac matrices only. Let us pick up terms proportional to $g_V^c$ in the above trace, which reads
\beqn &&{\rm tr} (P\!\!\!\!/+m_{J/\psi}){\epsilon\!\!/}_{J/\psi}[\gamma^\mu(P\!\!\!\!//2-q\!\!\!/+m_c)\gamma^\nu
 +\gamma^\nu(q\!\!\!/-P\!\!\!\!//2+m_c)\gamma^\mu]g_V^c\nonumber\\
 &=& {\rm tr} (P\!\!\!\!/+m_{J/\psi}){\epsilon\!\!/}_{J/\psi}[2 m_c g^{\mu\nu}+\gamma^\mu(P\!\!\!\!//2-q\!\!\!/)\gamma^\nu-
 \gamma^\nu(P\!\!\!\!//2-q\!\!\!/)\gamma^\mu]g_V^c.
\eeqn
It is seen that, after performing the trace, the second and third terms inside the above bracket will cancel each other, and the first term will give the contribution proportional to ${\epsilon}_{J/\psi}\cdot P$, which is equal to zero for the on-shell $J/\psi$ particle. Therefore $g_V^c$ will disappear in the amplitude of $J/\psi\to\gamma Z^*\to \gamma \nu\bar{\nu}$, and one can reach eq. (\ref{amplitude1}) by completing the trace in eq. (\ref{a2}) and converting $Z^*$ into the neutrino pair via eq. (\ref{weakneutralcurrent}). Also eqs. (\ref{approximation}) and (\ref{GF}) should be used in the derivation. The decay amplitude of Eq. (\ref{ampetac}) for $\eta_c\to \gamma\nu\bar{\nu}$  can be derived in the similar way. Due to the $\gamma_5$ appearing in eq. (\ref{projectoretac}), it is easily understood that $g_V^c$ other than $g_A^c$ will survive in the pseudoscalar charmonium case.

\end{document}